# Air-core photonic band-gap fibers: the impact of surface modes

**K. Saitoh**

*Hokkaido University, North 13 West 8, Kita-ku, Sapporo, 060-8628, Japan*
*ksaitoh@ice.eng.hokudai.ac.jp*

**N. A. Mortensen**

*Crystal Fibre A/S, Blokken 84, DK-3460 Birkerød, Denmark*
*nam@crystal-fibre.com*

**M. Koshiba**

*Hokkaido University, North 13 West 8, Kita-ku, Sapporo, 060-8628, Japan*
*koshiba@ice.eng.hokudai.ac.jp*

**Abstract:** We study the dispersion and leakage properties for the recently reported low-loss photonic band-gap fiber by Smith *et al*. [Nature **424**, 657 (2003)]. We find that surface modes have a significant impact on both the dispersion and leakage properties of the fundamental mode. Our dispersion results are in quantitative agreement with the dispersion profile reported recently by Ouzounov *et al*. [Science **301**, 1702 (2003)] though our results suggest that the observed long-wavelength anomalous dispersion is due to an avoided crossing (with surface modes) rather than band-bending caused by the photonic band-gap boundary of the cladding.

**References and links**

1. J. C. Knight, "Photonic crystal fibres," Nature **424**, 847-851 (2003).
2. J. C. Knight, J. Broeng, T. A. Birks, and P. S. J. Russell, "Photonic band gap guidance in optical fibers," Science **282**, 1476-1478 (1998).
3. R. F. Cregan, B. J. Mangan, J. C. Knight, T. A. Birks, P. S. J. Russell, P. J. Roberts, and D. C. Allan, "Single-mode photonic band gap guidance of light in air," Science **285**, 1537-1539 (1999).
4. C. M. Smith, N. Venkataraman, M. T. Gallagher, D. Müller, J. A. West, N. F. Borrelli, D. C. Allen, and K. W. Koch, "Low-loss hollow-core silica/air photonic band-gap fibre," Nature **424**, 657-659 (2003).
5. D. G. Ouzounov, F. R. Ahmad, D. Müller, N. Venkataraman, M. T. Gallagher, M. G. Thomas, J. Silcox, K. W. Koch, and A. L. Gaeta, "Generation of megawatt optical solitons in hollow-core photonic band-gap fibers," Science **301**, 1702-1704 (2003).
6. G. Bouwmans, F. Luan, J. C. Knight, P. S. J. Russel, L. Farr, B. J. Mangan, and H. Sabert, "Properties of a hollow-core photonic bandgap fiber at 850 nm wavelength," Opt. Express **11**, 1613-1620 (2003), http://www.opticsexpress.org/abstract.cfm?URI=OPEX-11-14-1613.
7. C. J. S. de Matos, J. R. Taylor, T. P. Hansen, K. P. Hansen, and J. Broeng, "All-fiber chirped pulse amplification using highly-dispersive air-core photonic bandgap fiber," Opt. Express **11**, 2832-2837 (2003), http://www.opticsexpress.org/abstract.cfm?URI=OPEX-11-22-2832.
8. N. A. Mortensen and M. D. Nielsen, "Modeling of realistic cladding structures for air-core photonic band-gap fibers," Opt. Lett. **29**, 349-351 (2004).
9. K. Saitoh and M. Koshiba, "Leakage loss and group velocity dispersion in air-core photonic bandgap fibers," Opt. Express **11**, 3100-3109 (2003), http://www.opticsexpress.org/abstract.cfm?URI=OPEX-11-23-3100.
10. J. Lægsgaard, N. A. Mortensen, J. Riishede, and A. Bjarklev, "Material effects in air-guiding photonic bandgap fibers," J. Opt. Soc. Am. B **20**, 2046-2051 (2003).
11. M.J. Steel, T.P. White, C. Martijn de Sterke, R.C. McPhedran, and L.C. Botten, "Symmetry and degeneracy in microstructured optical fibers," Opt. Lett. **26**, 488-490 (2001).

12. K. Saitoh and M. Koshiba, "Photonic bandgap fibers with high birefringence," IEEE Photon. Technol. Lett. **14**, 1291-1293 (2002).

## 1. Introduction

Silica-air micro-structured optical fibers are generally attracting a considerable interest because of their novel and unique optical properties (for recent reviews we refer to Ref. [1] and references therein) and the discovery of photonic band-gap (PBG) fibers [2, 3] has in particular strongly stimulated the strive for a low-loss PBG fiber with also a low dispersion and a high power threshold for non-linearities. Recently, a PBG fiber with a loss down to 13 dB/km at 1500 nm was achieved [4] and this fiber was also found to have very unusual dispersion properties [5].

From a modeling point-of-view the fiber structures have usually been considered ideal with circular air-holes but in order to model realistic PBG structures with large air-filling fractions one has to go beyond the circular representation. Experimentally, the cladding air-holes tend to be non-circular with a close-to-hexagonal shape at large air-filling fractions [4, 6, 7]. Recently, this was taken into account in a study of the PBGs in the cladding structures [8] and the results were found to be in excellent agreement with the cladding PBG reported in Ref. [4]. In this paper we extend the approach to a study of guided modes in the full fiber structure. We use a finite-element approach which allows us to calculate both dispersion and confinement properties [9] in a structure corresponding to that in Ref. [4], see Fig. 1. The role of confined surface modes has been debated and it has been suggested that a large part of the present loss level may be attributed to coupling of power to the surface modes [4]. Our results confirm the existence of surface modes and we also find that their presence strongly affects the dispersion properties of the fundamental mode via so-called avoided crossings of the photonic bands. We find finger prints of avoided crossings in the leakage loss as well as in the group-velocity dispersion (GVD) which agrees quantitatively with measurements [5].

## 2. Full-vector finite-element approach

We use the full-vector finite-element approach described in detail in Ref. [9] and references therein. Here, we briefly list the most important definitions. For a given frequency $\omega = ck = c2\pi/\lambda$ (where $c$ is velocity of light, and $k$ and $\lambda$ are the free-space wavenumber and wavelength, respectively) the numerical calculation provides us with a complex propagation constant $\gamma(\omega) = \beta(\omega) + i\alpha(\omega)$ where $\beta$ is the usual propagation constant of the plane-wave along the fiber axis and $\alpha$ is the attenuation constant associated with the exponential decay along the fiber axis. We present the photonic bands by their mode-index $\beta/k$ and the attenuation on a dB-scale by $20 \times \log_{10}(e) \times \alpha \cong 8.686 \times \alpha$. The group-velocity dispersion is expressed by the dispersion-parameter $D_w = \partial^2\beta/(\partial\lambda\partial\omega)$ where $\partial\omega/\partial\beta$ is the group velocity.

## 3. Dielectric structure

For the cladding structure we use the recently proposed two-parameter representation [8] shown in the lower left panel of Fig. 1, where the air-holes are hexagons (edge-to-edge distance $d$) with curved corners (curvature $d_c$). For a triangular arrangement the air-filling fraction is given by

$$f = \left(\frac{d}{\Lambda}\right)^2 \left[1 - \left(1 - \frac{\pi}{2\sqrt{3}}\right)\left(\frac{d_c}{d}\right)^2\right] \qquad (1)$$

where $\Lambda$ is the pitch. The hexagon-edges are tangents to circles defining the rounding of the corners so that the glass-air interface has no kinks and is smooth and well-defined.

The structure in Ref. [4] has $\Lambda \cong 4.7$ μm and $f \cong 0.94$ and by trial-and-error we find that $d/\Lambda = 0.98$ and $d_c/\Lambda = 0.44$ generates a structure which resembles the experimental structure surprisingly well, see Fig. 1. For the core we use the model shown in the lower middle panel of Fig. 1. Again, we start from a hexagon with rounded corners (curvature $D_c$), but in this case the hexagon-edges are not tangents to circles defining the rounding of the corners. To ensure a smooth and well-defined glass-air interface without unphysical kinks we introduce additional rounding (curvature $D_c'$) giving rise to small concave surface sections (when observed from the glass side of the interface). When this core structure is superimposed on the cladding structure (we consider the situation when the core is formed by removing 7 air holes from the cladding structure), six of the neighboring air-holes turn from hexagons to pentagons with rounded corners (curvatures $d_c$ and $d_c'$), see lower right panel in Fig. 1. By trial-and-error we find that $D = 14.0$ μm and $D_c = 0.59D$, $D_c' = 0.202D$, and $d_c'/\Lambda = 0.1$ results in a core region very similar to the real core in Ref. [4].

For the dielectric properties of the hybrid air-silica structure we use the refractive index $n = 1$ for the air regions and $n = 1.45$ for the silica regions. Neglecting the frequency dependence of the latter amounts to neglecting material dispersion which is somewhat justified for the fundamental mode which is mainly localized in the air-core region. Care must however be taken when the fraction of field-energy in silica increases [10].

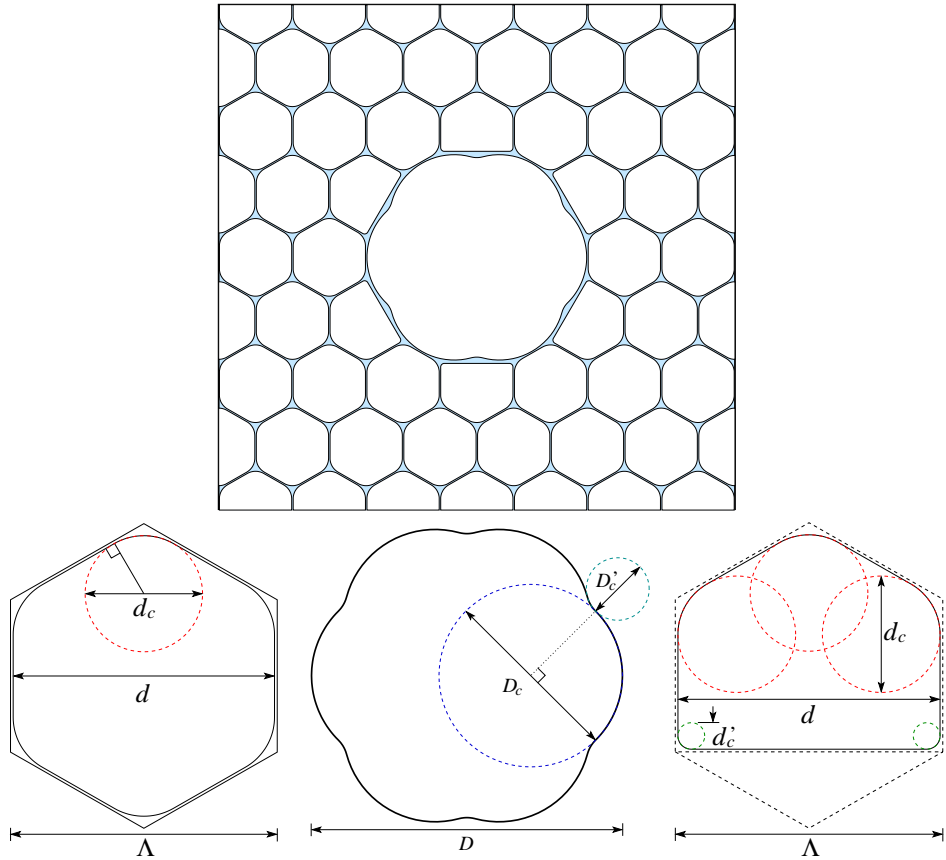

Fig. 1. The upper panel shows a dielectric structure that resembles that of Ref. [4]. The lower left panel shows a unit cell of the cladding region, the lower middle panel shows a close-up of the core construction, and the lower right panel shows a pentagon with rounded corners neighboring the air-core region.

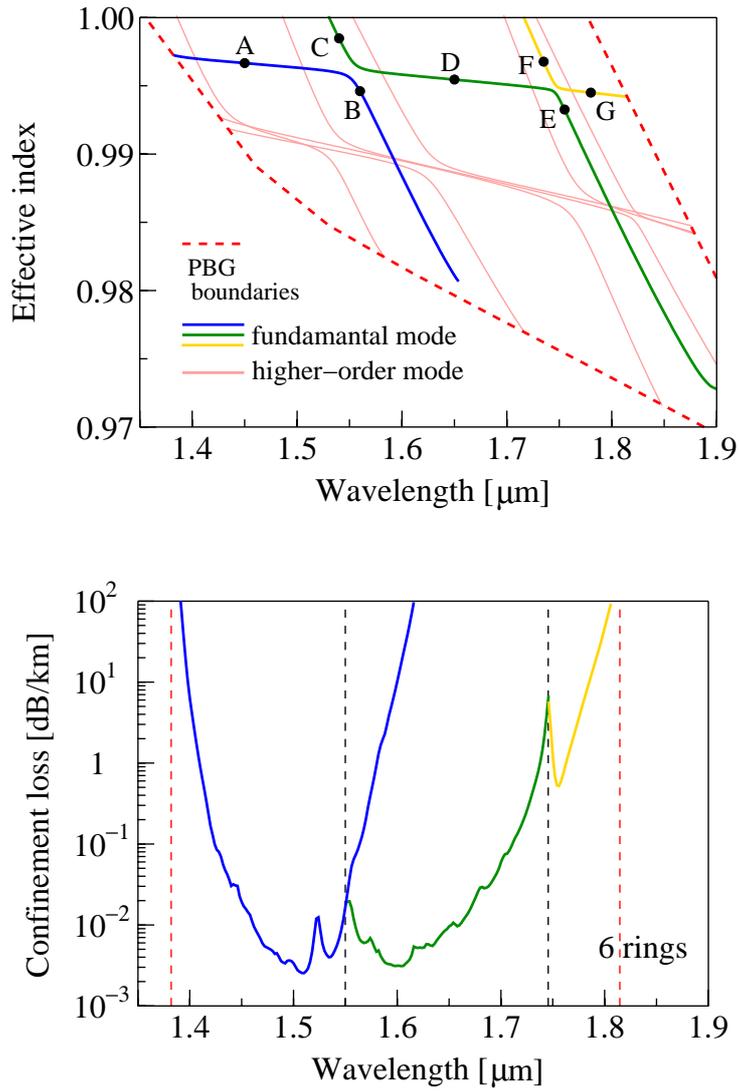

Fig. 2. In the upper panel the effective index versus wavelength plot illustrates the hybridization and avoided-crossings for guided air-core modes (low slope curves) and silica-surface modes (steep curves). Field plots around the avoided crossings are shown in Fig. 3. Lower panel shows confinement loss versus wavelength for the fundamental-like mode with PBG boundaries and avoided crossings indicated by dashed lines. The BPG boundaries are calculated for an infinite periodic lattice of the cladding holes.

## 4. Numerical results

In the upper panel of Fig. 2 we show the mode index of modes localized to the air-core region and the surrounding silica surface. We find that the cladding PBG confines two degenerated fundamental modes [11, 12] and four nearly degenerated high-order modes with a different polarization to the air-core region. In the mode-index representation all of these modes have photonic bands which are very flat as a consequence of a very small overlap with the silica. However, the PBG and the core shape also support surface modes. Since the surface modes are tightly confined to very small silica regions the corresponding photonic bands are rather steep (compared to the air-core modes) and having two different classes of modes (with steep and flat slopes) very often leads to hybridization and avoided crossings of the modes. In other words the eigenmodes of the system become linear combinations of the bare air-core and surface modes. In the mode-index plot the avoided crossings are clearly seen (note that in some cases crossings are allowed for symmetry reasons). In Fig. 3 we show contour plots of the modes at different points near the avoided crossings. As seen from panels A, D, and G the upper-most, almost flat, index curve corresponds to a fundamental-like mode mainly localized in the air-core region. In the same way we find that the other, almost flat, index curves correspond to high-order modes also confined to the core region (fields not shown). For the steep index curves panels B, C, E, and F clearly illustrate their surface-mode nature. As seen the surface modes are supported by the glass region surrounding the air-core and we believe that a smaller amount of glass near the core is a possible direction for suppressing the presence of surface modes.

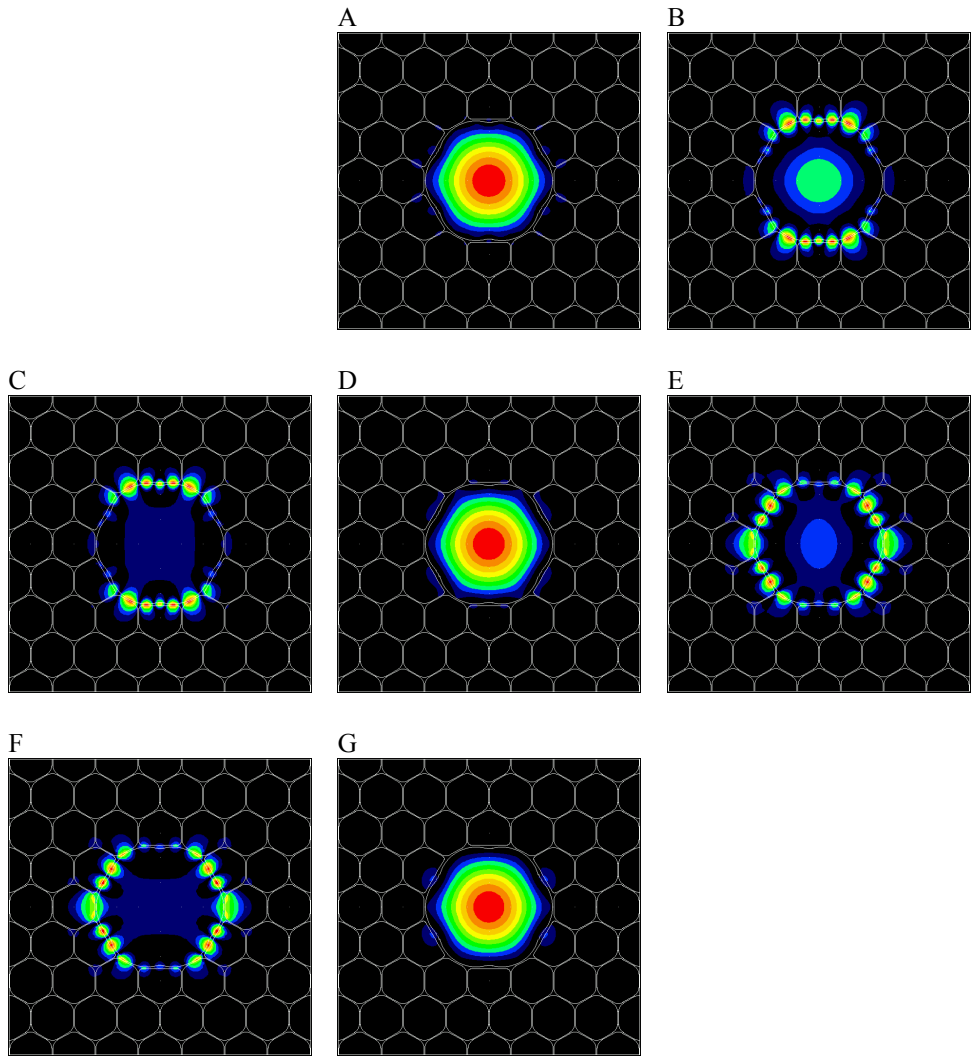

Fig. 3. Contour plots of the fundamental-like mode (panels A, D, and G) and surface modes (panels B, C, E, and F). The labeling of the panels refers to the labeling in Fig. 2.

In the lower panel of Fig. 2 we show the confinement loss for the fundamental-like mode in the case of a structure with 6 rings of air holes. As seen, the attenuation spectrum carries strong finger prints of the avoided crossings (indicated by black dashed lines) and similar finger prints can be expected in a scattering loss spectrum because the beat-length between the fundamental mode and the cladding and/or surface modes also carries information about the avoided crossings.

We also note that the confinement loss increases toward the high-wavelength PBG boundary which agrees with the experimental observation in Ref. [4], though in experiments there are additional contributions from scattering loss and coupling of power to the surface modes. In general confinement loss of course increases in the vicinity of PBG edges. However, the strong increase in confinement loss upon hybridization with the surface modes also indicates that the bare surface modes are leakier than the bare air-core modes. We believe the origin is at least two-fold: i) the surface modes are more tightly confined than the air-core modes which results in a somewhat higher numerical aperture (like high-order air-core modes, their Fourier transforms have large transverse k-components) and ii) the surface modes are localized closer to the homogeneous glass region surrounding the PBG cladding.

In Fig. 4 we show the GVD for the fundamental mode which has zero dispersion around $\lambda \cong 1.450$ μm in reasonable agreement with the experimental result $\lambda \sim 1.425$ μm [5]. The GVD is normal at the short-wavelength PBG boundary and toward the long-wavelength regime it becomes anomalous and in Ref. [5] this behavior was suggested to be a typical Kramers-Kronig-like phenomena. In our simulations we find that the anomalous dispersion arises as a consequence of an avoided crossing with a surface mode rather than the long-wavelength PBG boundary of the cladding. It should be noted that in the experiments in Ref. [5] it is quite likely that some linear combination of hybrid air-core and surface modes is excited rather than just the fundamental air-core mode. This makes the interpretation difficult and a direct quantitative comparison to our calculation of the fundamental eigenmode GVD is only reasonable when the wavelength is not too close to neither the PBG edges nor the avoided-crossings.

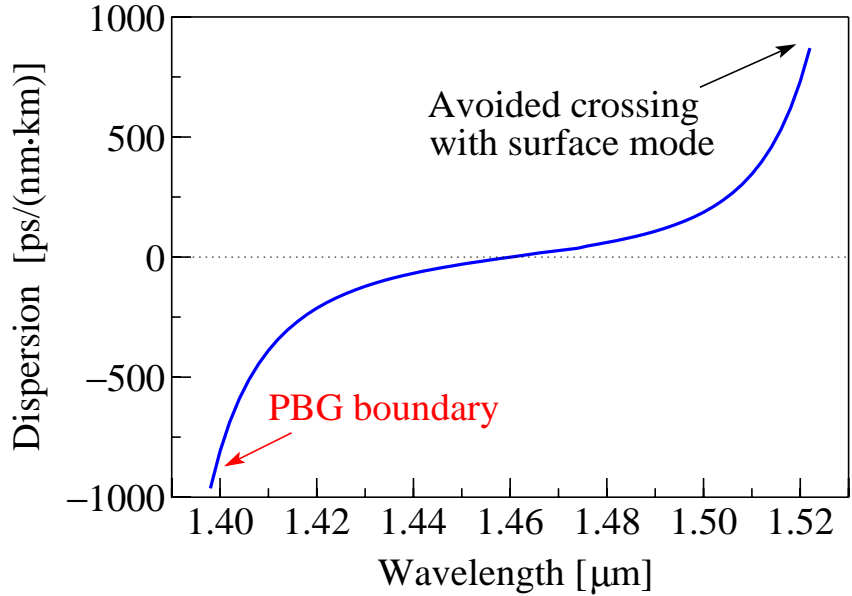

Fig. 4. Plots of the group-velocity dispersion versus wavelength for the fundamental-like mode, see Fig. 2 and panel A in Fig. 3.

## 5. Conclusion

We have studied the dispersion and leakage properties of the PBG fiber in Ref. [4, 5] by aid of a finite-element approach [9] and the recently proposed two-parameter representation for the air-holes in the cladding [8] as well as for the air-core. Our simulations support the existence of surface modes as well as high-order modes in the air-core. We have studied the dispersion and leakage properties of the fundamental mode and found that the surface modes have a strong impact on these properties. Our results also illustrate how the presence of surface modes prevents broad-band transmission (over the full PBG of the cladding) with low group-velocity dispersion. On the other hand, the surface modes facilitate strong anomalous dispersion (via avoided crossings) which may find use in novel applications [5, 7]. While our quantitative observations may not be universal we believe that the qualitative behavior is quite general; when surface modes are present it is very likely that they strongly modify the dispersion, leakage loss, and scattering loss of the bare air-core modes.

## Acknowledgements

N. A. Mortensen acknowledges stimulating discussions with M. D. Nielsen and A. L. Gaeta.